\documentstyle[12pt,epsf,psfig]{article}

\def\be{\begin{equation}}
\def\bea{\begin{eqnarray}}
\def\ee{\end{equation}}
\def\eea{\end{eqnarray}}

\def\ra{\rangle}
\def\la{\langle}

\def\D{\Delta}
\def\eps{\epsilon}

\def\th{\theta}

\begin{document}

\title{Phase Transitions of Single Semi-stiff Polymer Chains}
\author{Ugo Bastolla and Peter Grassberger\\
{\sl \small HLRZ, Forschungszentrum J\"ulich, D-52425 J\"ulich, Germany}}
\maketitle

\medskip
\medskip

\begin{abstract}
We study numerically a lattice model of semiflexible homopolymers 
with nearest neighbor attraction and energetic preference for 
straight joints between bonded monomers. For this we 
use a new Monte Carlo algorithm, the `Pruned-Enriched 
Rosenbluth Method' (PERM). It is very efficient both 
for relatively open configurations at high temperatures, and
for compact and frozen-in low-$T$ states. This allows us to 
study in detail the phase diagram as a function of nn 
attraction $\epsilon$ and stiffness $x$. It shows a $\th$-collapse 
line with a transition from open coils (small $\epsilon$) to molten 
compact globules (large $\epsilon$), and a freezing transition
toward a state with orientational global order (large stiffness $x$).
Qualitatively this is similar to a recently studied mean field theory 
(S. Doniach, T. Garel and H. Orland (1996), {\it J. Chem. Phys. \bf
  105} (4), 1601), but there are important differences in details. 
In contrast to the mean field theory and to naive expectations, the 
$\th$-temperature {\it increases} with stiffness $x$. The freezing 
temperature increases even faster, and reaches the $\th$-line at 
a finite value of $x$. For even stiffer chains, the freezing 
transition takes place directly, without the formation of an 
intermediate globular state. Although being in conflict with 
mean field theory, the latter had been conjectured already by 
Doniach {\it et al.} on the basis of heuristic arguments and of 
low statistics Monte Carlo simulations.

Finally, we discuss the relevance of the present model as a very 
crude model for protein folding.
\end{abstract}

\centerline{Keywords: Polymers, Protein Folding, Phase Transitions}

\section{Introduction}
The statistical mechanical study of protein folding \cite{PF} is still 
at its beginning. Minimal models try to represent its gross features 
by incorporating only those few ingredients that are
supposed basic for its qualitative understanding. Mainly with 
this motivation, Doniach {\it et al.} \cite{DGO} studied recently 
a model of semi-stiff lattice chains. In this model monomers 
are located at the sites of a simple cubic lattice. An attraction 
between non-bonded nearest neighbors was included to mimic the effect 
of average hydrophobicity. In order to induce an ordering phase 
transition between a random (molten) globule and a frozen configuration 
-- which would mimic a uniquely folded protein --, an interaction was 
included which favored straight joints between bonds along the polymer 
backbone over rectangular joints. One way to interpret this, pointed 
out in \cite{DGO}, is to interpret each ``monomer" as a $\alpha$-helical 
turn (ca. 3 amino acids), and to consider the ordering transition as 
a transition to a protein consisting only of $\alpha$-helices. 

Depending on the temperature and the chain stiffness, three phases 
were indeed found in \cite{DGO} by means of a mean field theory: 
an open coil at high $T$, a collapsed but `molten' globule at 
intermediate $T$ and low stiffness, and a `frozen' state at low $T$ 
and large stiffness. The coil-globule transition had all typical 
features of the $\th$-transition found at zero stiffness: it is 
second order (indeed, it is a tricritical phenomenon \cite{DG}), 
and $T_\th$ should depend only weakly on chain stiffness. The 
freezing transition at $T=T_F$ should however be first order. Although 
the mean field theory predicted that $T_F<T_\th$ for all values of 
the stiffness, it was conjectured in \cite{DGO} that this might 
indeed not be correct, and that $T_F$ might indeed become larger 
than $T_\th$ for sufficiently stiff chains. If that is the case, 
then one should observe a direct first order transition from open 
coils to ordered states for very stiff chains. This was at least 
not contradicted by Monte Carlo simulations made by the same authors 
\cite{DGO}, but the simulations were hardly convincing as the 
authors were not able to simulate sufficiently long and stiff chains.

Actually, models similar to the above had been studied already much 
earlier \cite{Fl,YB,BY,Baum,Kol,Kol2,Mans} as models for other 
semi-stiff polymers like, e.g., DNA. Baumg\"artner {\it et al.} and 
Mansfield \cite{YB,BY,Baum,Mans} 
studied indeed melts consisting of many short semi-stiff chains. Thus 
they could not address the problem of $\th$-collapse, but they 
showed very convincingly that there is a first-order ordering 
transition on the simple cubic lattice, while there is presumably 
a second order ordering transition in 2-d. Since the transition in 3-d 
was found for long chains and did not seem to become smoother with 
chain length, it is very plausible that it should coincide with 
the freezing transition found in \cite{DGO}, in the limit of infinite 
chain length. The results of \cite{DGO} and \cite{YB,BY,Baum,Mans} are 
indeed fully consistent. 

The opposite case of a single chain, but at parameters where freezing 
was out of range, was studied in \cite{Kol,Kol2}. These authors 
were mainly interested in the problem whether stiffness increases 
or decreases the theta temperature $T_\th$. Naively one might 
expect that stiffness should make collapse less easy, and should 
therefore decrease $T_\th$. A swelling with increased stiffness at fixed 
$T\approx T_\th|_mbox{non-stiff}$ was indeed seen in in \cite{Kol,Kol2} 
for short chains. But these authors were careful to point out that this
might be a finite-size effect, and that the actual value of $T_\th$ 
(defined via the limit $N\to\infty$) might actually increase. For very 
long chains, stiffness might actually foster collapse: once a hairpin 
has been made, it might be much easier to follow a stiff chain
than a completely flexible one. As we said already, the mean field 
theory of \cite{DGO} predicted that this effect should invalidate
the naive argument, and $T_\th$ should be independent of stiffness.

Seen as a rudimentary protein model, the above model lacks of 
course one essential ingredient, namely heterogeneity. It is usually 
assumed that heterogeneity between individual amino acids is the 
main force which drives a collapsed polypeptide into a unique 
native configuration. Nevertheless, it might be that stiffness 
plays a similar role as heterogeneity, in which case the model 
of \cite{DGO} might catch typical features of real protein folding. 
Indeed, in a recent treatment of random copolymers \cite{MKD} the 
authors found a phase diagram (fig.3 of \cite{MKD}) which is 
surprisingly similar to the one found in \cite{DGO}. 

To elucidate these intriguing questions, we decided to perform 
more extensive Monte Carlo simulations. A further motivation was 
to test a novel algorithm, the Pruned-Enriched-Rosenbluth-Method (PERM) 
developed by one of us \cite{PERM}. This is a chain growth algorithm 
superficially similar to the one used also in \cite{DGO}, but with 
some essential differences. It has proven extremely efficient in a 
number of problems, most of which involved however rather open 
configurations: free SAW's \cite{Gr97}, $\th$-collapse of flexible 
chains \cite{PERM}, and coagulation transition in dilute solutions 
\cite{FG}, just to name a few. We wanted to see how it performs at 
very low energies and near first order phase transitions, before 
using it in more realistic studies of protein folding. 
 
The paper is organized as follows. After a brief description of the
model and of the algorithm that we used (section 2), we present our
numerical results concerning the transition to the globular phase
(section 3) and the freezing transition (section 4). In section 5 we
finally discuss these results, draw our conclusions, and point out
further open problems.

\section{The Model and the Algorithm}

We represent a polymer as a self-avoiding random walk (SAW) on a 
simple cubic lattice \cite{DG}. Thus, monomers are placed on the 
lattice sites, and double occupancy of a site is strictly forbidden. 
Boundary conditions will be discussed below, as we used different 
ones for different purposes.
The energy of the chain takes into account two contributions: a 
negative energy $-\eps$ for each non-bonded occupied nearest-neighbor 
pair (this is the standard attraction used in simulations of 
$\th$ polymers \cite{HG}), and a positive energy $x\eps$ for each 
change of direction of the walk, i.e. for each non-collinear pair 
of successive bonds. The parameter $x$ will be called stiffness. 
Depending on the interpretation of the model, it might represent 
the fact that {\it trans} bonds are energetically favored over 
{\it gauche} bonds, or that $\alpha$ helices are favored over random 
secondary structures.

In the following we shall assume that $\eps=k_B$, or in other words
we shall measure temperatures in units of $\eps /k_B$ where $k_B$ is 
the Boltzmann constant. We will also use sometimes the Boltzmann factor 
$q=e^{\eps /k_B T}=e^{1/T}$ as a control parameter.

The algorithm that we use, PERM, is described in detail in \cite{PERM}. 
Here we recall for completeness its main aspects, adding some 
technical details which were important to simulate systems at very 
low temperature.

The starting point of PERM is the Rosenbluth-Rosenbluth method (RR),
developed already in 1955 \cite{RR}. In this method, a chain is built 
by adding a new monomer at each time step. Assume we are at step $n$, 
and the last placed monomer has $m_{n-1}$ free neighbors. If $m_{n-1}\geq 1$, 
the new monomer is placed with some probability $p_n(k)$ in the 
$k$-th free neighbor site, and the algorithm continues. If not, we 
discard the chain and start a new one. Irrespective of the precise 
form of $p_n(k)$ this would introduce a bias towards compact 
configurations with few free neighbors, if all generated chains were 
given the same weight. Thus each generated configuration carries a weight
which compensates for this bias. In addition, this weight will take
care of the Boltzmann weight. In the simplest case of uniform neighbor 
sampling, $p_n(k)=1/m_{n-1}$, the total weight of a chain should be
\be 
   W_N = \prod_{n=1}^N w_n(k_n) 
\ee
with
\be 
   w_n(k)=z m_{n-1}e^{-E_k/k_B T}, 
\ee
$z$ being a fugacity we are free to choose.
More generally, we can use any $p_n(k)$ (provided it is non-zero for 
each allowed neighbor), and weights 
\be 
   w_n(k)=z {e^{-E_k/k_B T} \over p_n(k)}.
\ee

This algorithm will be optimal if we manage to keep $w_n(k)$ constant 
and to avoid traps in which $m_n=0$: in this ideal case -- which is 
of course impossible to reach in practice -- each attempted chain 
growth would succeed, and we would have perfect importance sampling.

In real life each factor $w_n(k)$ will fluctuate, giving roughly a 
lognormal distribution for $W_N$. Thus, for very long chains, the RR 
ensemble is dominated by rare configurations with very high weight,
leading to serious statistical problems \cite{BK}.

To overcome this difficulty, PERM uses another classical idea of 
polymer simulations: enrichment \cite{enr}. Originally, this was 
devised as a method to overcome attrition, i.e. the exponential 
decrease of the number of successful attempts in `simple sampling' 
(here, in contrast to the RR method, {\it all} neighbors of the 
last placed monomers are sampled with the same probability, whether 
they are free or not). It consists simply in replacing unsuccessful 
attempts by copies of successful ones (in this respect enrichment is 
similar to a genetic algorithm). In PERM, enrichment is implemented 
by monitoring the weight $W_n$ of partially grown chains. If 
$W_n$ exceeds some preselected upper threshold $W_n^>$, we make 
two or more copies of the chain; divide $W_n$ by the number of 
copies made; place all except one onto a stack; and continue with the 
last copy. In this way the total weight $W_n$ is exactly preserved, 
but it is more evenly spread on several configurations. This is 
done at every chain length $n$. 

The last entry to the stack is fetched if the current chain has 
reached its maximal length $N$, or if we ``prune" it. Pruning (the 
opposite of enrichment) is done when the current weight $W_n$ has 
dropped below some lower threshold $W_n^<$. If this happens, we 
draw a random binary number $r_n$ with ${\rm prob}\{r_n=0\} = 
{\rm prob}\{r_n=1\} = 1/2$. If $r_n=1$, we keep the chain but double 
its weight. If not, we discard (``prune") it, and continue with the 
last entry on the stack. If the stack is empty, we start a new chain.
When the latter happens, we say we start a new ``tour". Chains within 
one tour are correlated, but chains from different tours are strictly 
uncorrelated except through the dependence of $W_n^<$ and $W_n^>$ on 
previous tours. 

One can easily see that this algorithm is correct in the sense that 
the average partition sum estimate agrees exactly with the exact 
partition sum, provided we estimate it only between finished tours 
(i.e. at empty stack), and provided the total number of tours itself 
is uncorrelated with the partition sum. The latter would not be the 
case if we would stop the algorithm after some preset CPU time (we 
would not sample properly very large tours, and put too much weight on 
small ones). We should also remember that the algorithm cannot 
give unbiased estimates for free energies and for observables which 
are essentially based on free energies (such as end-to-end distances, 
which are derivatives of $F$ with respect to some external field), 
if the fluctuations of the partition sum estimates are large. This is 
the main problem at very low temperatures.

In the code that we use, the algorithm is implemented recursively. A 
subroutine adds a monomer at a time. When we want to enrich a chain, we
call the subroutine twice (or more, if we want to place more than one 
copy onto the stack), reducing accordingly the weight. When we want to 
prune a chain, we leave the subroutine without calling itself with 
probability 1/2, and else double the weight. Actually, instead of 
placing copies onto the stack at each enrichment event, we keep only 
one master copy which is updated at each step. In addition, we have 
an array of counters which tells us for every chain length how many 
copies of this length are still to be handled, and an array which 
shows the occupancy of each lattice site. 

This implementation is in contrast to the implementation of the 
enrichment idea in \cite{DGO}, which is more in spirit of a 
genetic algorithm: at each time, a large population of copies is 
kept in memory, each of the same length $n$, and pruning and 
enrichment are done by replacing low-weight chains by copies of 
high-weight chains. This can be efficient on large parallel 
machines, but it poses formidable storage and data transfer problems.
For finite-size populations there are also corrections which are 
not easy to analyze.

A crucial advantage of PERM is the fact that all controls -- the 
selection probability $p_n(k)$, the thresholds $W_n^>$ and $W_n^<$, 
and the number of copies made -- are dummies which can be modified 
at {\it any} stage of the run. We can thus adjust them automatically 
in response to problems which might arise. This is a major improvement 
over other recursive chain growth methods \cite{HG,NB}, where 
the fugacity is a control parameter which cannot be changed during 
the run without introducing a bias.

A first good choice is to take $W^>_n$ and $W^<_n$ proportional to 
the current average of $W_n$, $W^>_n = c_+\la W_n\ra$ and 
$W^<_n = c_-\la W_n\ra$, with coefficients $c_\pm = O(1)$ \cite{PERM}. 
Notice that with this choice the dependence 
on the fugacity $z$ drops out, and we can choose it arbitrarily.
For temperatures close to $T_\th$ we used this with $c_+ = 3$ to 10, 
and $c_+/c_- \approx 10$ to 50. Within this wide range of parameters 
this was very efficient. 

Problems arose however at very low $T$, since there the current 
estimate $\la W_n\ra$ might be very wrong. The most dangerous and common 
situation is that we underestimate $\la W_n\ra$ since we have not yet 
encountered a ``good" (i.e., low energy) configuration. When 
such a configuration finally appears, our thresholds are too low 
(occasionally by many orders of magnitude), and we produce an 
enormous amount of copies without pruning them sufficiently. Our way 
out of this dilemma consists in letting $c_\pm$ depend on the  
total number of copies of length $n$ already made during the present 
tour. If this number is becoming too large, we reduce $c_\pm$. But this 
alone would reduce the total number of very long chains produced, since 
these are most effected by fluctuations. The algorithm is most efficient 
if the sample size is the same for all chain lengths. We cannot enforce 
this precisely, but we can place a strong bias towards it by 
choosing $c_\pm$ proportional to the total sample size. For technical 
reasons, we indeed segmented the set of tours into bunches of typical 
size $10^2$ to $10^3$. We replaced the number of configurations made 
during the present tour by that made during the present bunch. Let us call 
this $M_n^{\rm bunch}$, while $M_n^{\rm total}$ is the total number 
of configurations generated so far. We then used 
\be
   c_\pm = c_\pm^0 \left({M_n^{\rm total} \over M_1^{\rm total}} 
        \right)^\alpha \left(1+M_n^{\rm bunch} / a\right)^\beta
\ee
with $c_+^0 \approx 5, c_+^0/c_-^0 = \approx 20$, $a\approx 10^3\;-\;
10^4$, and exponents $\alpha$ and $\beta$ either 1 or 2. The latter 
was needed for the lowest values of $T$.

The prefactors $c_\pm^0$ can be learned from preliminary runs with 
small chain lengths. A more systematic 
strategy which was quite successful consisted in a rudimentary 
genetic algorithm (with mutation and replacement by the fittest only, 
but without cross-over) by which we let a population of pairs 
$(c_+^0,c_-^0)$ evolve.

At high temperatures, it was sufficient to make just one new copy per 
enrichment event. At low $T$ this was not enough to prevent weights 
from growing too large. We used a number of schemes which all allowed 
a large number of copies, but only if $W_n$ grew excessively large. A 
good choice was $n_{\rm copy} = {\rm int}[\sqrt{W_n/W_n^>}]$.

The last point which can change the efficiency of the algorithm is 
the neighbor selection probability $p_n(k)$. In agreement with 
\cite{PERM} we found that it was not useful to include in it the 
Boltzmann weight for the contact energy. Most likely, a large number 
of contacts gives an immediate advantage but leads to a higher risk 
of getting blocked later. But for large $x$ we found it important to 
favor straight steps over right angles. In the simplest case we thus 
used 
\be 
   p_n(k) = \left\{\begin{array}{c@{\qquad}l}
           1/m_{n-1} & \mbox{going straight is blocked} \\
           e^{\beta x}/(m_{n-1} + e^{\beta x}-1) & \mbox{going straight} \\
           1/(m_{n-1} + e^{\beta x}-1) & \mbox{right angle turn, although 
                    straight is not blocked} \end{array} \right. 
\ee
In the ordered phase, it was useful to go even further and enhance 
$p_n(k)$ even more for straight steps if the previous step had also 
been straight.

\section{The $\th$-collapse line}

For small stiffness, the end-to-end distance $R_e$ is the easiest 
and most straightforward indicator for the coil-globule transition. 
At $T_\th$, we expect $R_{e,N}\sim N^{1/2}$ up to logarithmic 
corrections \cite{DG,HG}. But, as we noted already, this collapse is 
delayed for stiff chains. Thus $R_e$ is not very useful for large 
values of $x$. This is illustrated in fig.1, where we see that we 
need chains of length $N>1000$ to pin down $T_\th$.

A more useful observable could be the second virial coefficient 
which should vanish at $T_\th$. But measuring it for interacting 
chains is a bit awkward \cite{HG}. For a-thermal chains there exists 
the very elegant Karp-Luby algorithm \cite{sokal}, but no similarly 
elegant algorithm seems to be known for thermal chains.

We found the most efficient method to measure $T_\th$ to be one due 
to Dickman \cite{dickman}. Here we use the fact that the pressure 
vanishes at $T_\th$, and that pressure is defined as the derivative 
of the free energy with respect to volume. We thus study 
simultaneously two systems in volumes of different sizes, 
and compare their partition functions. More precisely, in order to 
reduce statistical fluctuations, we start with a single chain on
a large lattice with periodic b.c. with period $L_1$ (actually we 
used `helical' b.c. where each site is labeled by a single index $i$, 
and $i+L_1^3 \equiv i$). At each monomer insertion we check that 
it would still be self avoiding and would not change its contact 
energy if we would replace the periodicity by $L_0 = L_1/2$. As 
soon as this is violated, we replace the chain by two copies, one 
on volume $L_1^3$ and the other on $L_0^3$, and measure the
quantity
\be 
   \D Z(N,L_0)={Z_N(L_0)-Z_N(L_1)\over Z_N(L_1)} 
\ee
where $Z_N$ is the partition sum.
This vanishes exactly for short chains. For very long chains, 
it must go to $-1$. For $T>T_\th$ it is negative for all $N$ 
(pressure is always positive), but for $T<T_\th$ there should exist 
a range where pressure is negative and thus $\D Z(N,L_0)>0$. For $x=10$ 
this is shown in fig.2. In this figure, the same set of temperatures 
are used as in fig.1. We see very clearly that all curves except one 
correspond to already collapsed chains. Actually, the argument is 
a bit more subtle due to finite size corrections which we have so 
far neglected. Simulations show that $\D Z(N,L_0)>0$ even at $T=T_\th$, 
but its maximum with respect to $N$ does not increase with $L_0$ unless 
$T<T_\th$. Thus a 
precise determination of $T_\th$ is possible by comparing different 
lattice sizes. Since $\max_N \D Z(N,L_0)$ is slowly varying with 
stiffness $x$, it is sufficient to make this finite size analysis at  
few values of $x$. 

The results are shown in fig.3. They fully agree with those obtained 
from $R_{e,N}$, but are much more precise for large $x$. Most of 
them were obtained with $L_0=2^6$ and $N\approx 5000$. The error on 
$T_\th$ is typically less than 1\%, independent of $x$. We see clearly 
that $T_\th$ increases with stiffness, i.e. stiffness {\it increases} 
the tendency to collapse.

\section{The freezing transition}

At low temperature the chains are expected to undergo a first order 
transition toward a phase characterized by global order, at an 
$x$-dependent temperature $T_F(x)$. The low energy
configurations should appear as a bundle of linear parallel pieces in 
contact one to each other, with as few turns as possible. 

To monitor the freezing transition, and to verify that it is first order, 
we observed four different quantities:

\begin{enumerate}
\item The average number of contacts per monomer, $\la m\ra$, which
  measures the contact energy. For large $N$ this quantity should 
  show a a discontinuity at $T_F$. But this discontinuity is very 
  broad at finite chain lengths, except at very large values of $x$. 
  Thus we have very large finite size effects, and $\la m\ra$ is not 
  practically useful as an order parameter.
\item The fraction $f_s$ of straight (trans) bonds, $f_s$. This measures 
  the stiffness energy and the local ordering of the chain. In 
  contrast to $\la m\ra$, this seems very useful as order parameter. 
  Above $T_F$ it is found to be close to the naive expectation 
  obtained for chains without self-avoidance and nn-attraction, 
  \be
     f_s={1\over (1+({\cal N}-2)q^{-x})}
  \ee
  (here ${\cal N}$ is the lattice coordination number; in our case,
  ${\cal N}=6$). This is a particularly good approximation for large 
  $x$. Deviations for small $x$ can largely be understood as effects 
  of self-avoidance. In the frozen phase $f_s$ is close to 1. 
\item The fraction of bonds directed in the privileged direction. Let 
  us denote by $n_x, n_y$, and $n_z$ the bonds parallel to either of 
  the three coordinate axes. Let us define $n_{\rm max}=\max_{i=x,y,z} 
  n_i$, $n_{\rm min}=\min_{i=x,y,z}n_i$, and $p=1-n_{\rm min}/n_{\rm max}$. 
  If there is no directional ordering at all, we have $\la p\ra \sim
  1/\sqrt{N}$ due to the central limit theorem. In the opposite case 
  of an ordered phase, we have $p\to const$ for $N\to\infty$. In the 
  intermediate case of weak directional ordering, a mean field type 
  argument predicts a power law decay, $p\sim N^{-\alpha}$, with 
  non-universal exponent $\alpha$. 
\end{enumerate}

Notice that neither $R_e$ nor the gyration radius are very useful for 
detecting the ordered phase. We expect to see some changes at $T_F$ 
since configurations should change from spherical globules to 
more rugged shapes, but we cannot expect this to be very systematic 
and easy to observe. For this reason, chain sizes were not measured 
for $T\approx T_F$, except for very stiff chains where $T_F$ is rather 
large and where finally the collapse is without intermediate globule 
state.

We found a dramatic dependence on $x$ in the ease of locating $T_F$. 
In contrast to \cite{DGO} who were not able to go beyond $x=3$, we 
found that the freezing transition is much easier to observe for large 
$x$ than for small $x$. This should have been expected intuitively: when 
cooling down at small $x$, we have first a collapse to a disordered 
globule, and ordering sets in only at very low $T$ when the mobility 
of the chain is extremely low. Thus we are bound to find important 
metastable states and long trapping times. For large $x$, in contrast, 
the ordering sets in at rather high $T$ when the chain is still highly 
mobile. To illustrate this, we compare in figs. 4 and 5 the behavior 
of $f_s$ for $x=3$ with that for $x=10$. In both cases we see quite 
clear phase transitions, but it is much sharper for $x=10$ than for 
$x=3$. The same is true for $p$, see figs. 6 and 7. 

Indeed, we were 
not able to obtain reliable results for $x=1$ and $N>150$, where 
the authors of \cite{DGO} claimed to see a clean ordering transition. 
The problem is that partition sum estimates fluctuate wildy in this 
region. At the freezing point for for $x=1$ and $N=200$, this could 
involve many orders of magnitude even in samples of several million 
chains. Thus even very large statistical samples were dominated by 
only few large-weight configurations. The authors of \cite{DGO} grew 
populations of only 20,000 chains. Our guess is that they grew many 
such populations for each set of control parameters, and averaged 
the results {\it without} weighting them with $Z_N$. Correct averages
should include this factor. Neglecting this would reduce greatly the 
statistical error estimates, at the risk of making uncontrolled 
systematic errors.

To locate precisely a first order transition, we should in principle 
make a finite size scaling analysis. We indeed see in figs. 4 to 7 
important finite size effects: with increasing $N$ the transition point 
seems to shift towards higher values of $T$. This is as 
expected: for small $N$ we have important surface effects which diminish 
the cooperativity of the interaction. But fluctuations are too large to 
allow a systematic analysis. 

In spite of all these problems, we were able to determine $T_F$ in a 
wide range of $x$ values (fig.3). Our estimates for $T_F$ for small $x$ 
agree nicely with those of \cite{DGO} and of \cite{YB,BY,Baum,Mans}. 
The ordering transition for concentrated chains (hamiltonian walks) 
\cite{YB,BY,Baum,Mans} should coincide with the presently studied 
transition in the limit $T\to 0$. From these references we expect 
$T_F/x \to 0.66$ \cite{BY} for $T_F\to 0$ resp. $T_F/x \to 0.82$ 
\cite{Mans} for $T_F\to 0$. Our data are in better agreement with the 
latter, although the large statistical errors for and the evident 
curvature of the transition line makes an extrapolation difficult.
A possible alternative fit to our data is 
$T_F \propto x^{0.65}$. This fit is indeed better numerically and 
holds for the entire range of $x$, but we see no theoretical basis 
for it at small values of $x$, and it would contradict all previous 
numerical \cite{YB,BY,Baum,Mans,DGO} and theoretical \cite{DGO} 
results.

The most conspicuous result is that $T_F$ reaches the coil-globule 
transition temperature at $x\approx 13$. Beyond this triple point, we 
have a direct first order collapse from open coils to folded structures. 
The existence of this direct transition is also seen in $R_e$ which 
drops suddenly at $T_F$ when $x>13$.

We stopped our simulations at $x>18$, but $T_F$ seems to continue to 
grow with $x$, and we see no good theoretical argument why it should 
not do so. Thus we conjecture that $T_F$ would finally tend to infinity. 

As we said already, the average total energy was not very informative. 
More interesting was the contact energy per monomer $\langle m\rangle
= \langle \#(\mbox{nn pair contacts})\rangle /N$.
Instead of showing it as a function of $N$ for various values of $(x,T)$, 
we present it in fig.8 as a function of $x$, at fixed $N$ and $T$. We 
expect two clear situations, one at $T>T_{\rm triple}$ and the other 
at $T<<T_\theta(0)$.
\begin{itemize}
\item In the high-$T$ phase, $T>T_{\rm triple}$, we expect that  
$\langle m\rangle$ decreases with $x$, until the freezing line is 
reached. At this points $\langle m\rangle$ should have a 
discontinuity for $N=\infty$. For finite $N$ this 
jump should be smeared, but it should still be rather sharp, and it 
should not show any precursor. As long as the chain is in the coil 
phase and not close to the $\theta$-line, increasing stiffness should 
decrease the number of contacts.
\item When $T$ is lower than $T_\theta(0)$, it is also lower than 
the collapse temperature for stiffness $x>0$. In this case we are from 
the beginning in the collapsed phase. When $T<<T_\theta(0)$, the 
density is rather high for all $x$, and we do not expect 
$\langle m\rangle$ to decrease 
initially. Instead, we expect first a weak increase with $x$, which 
accelerates when we pass through the freezing line. This time the 
freezing line is however much less sharp, and we expect much stronger 
finite-$N$ corrections.
\end{itemize}
Both predictions are supported by fig.8. Most impressive is the 
decrease and subsequent jump for large $T$. The behavior at low 
temperatures is less clear, as there are still strong small-$N$ 
corrections. In particular, there is an initial decrease with $x$ 
for small $N$. Most difficult to interpret are data at 
intermediate $T$. The data for $T=3.811$ cross the $\theta$-line 
at $x\approx 1$. For $N=\infty$ we should have an increase there 
with infinite slope, as the specific heat diverges logarithmically at 
$T_\theta$. But the data show a steady and systematic decrease. Obviously 
this is a finite-$N$ effect which dominates completely the behavior up 
to extremely large values of $N$. When we approach the folding line
$\langle m\rangle$ finally increases, with an $N$-dependence 
intermediate between the two previous cases.

\section{Discussion}

We have been able to map out a large region of the phase diagram for 
semi-stiff chains with self avoidance and nearest-neighbor attraction. 
This region contains the coil-globule transition (second order), the 
freezing transition from molten globule to a folded state (first order), 
and a triple point where these transition lines meet. We found our 
algorithm very efficient, in particular for very stiff chains where we 
had expected the biggest problems when we started this investigation.

In our study we observed three features of the phase diagram which
were not expected on the ground of mean field theory, although the
most important one, {\it i.e.} the existence of a triple point, was
already conjectured in \cite{DGO}:

\begin{enumerate}
\item The collapse temperature $T_\th(x)$ is an increasing function of the
  stiffness;
\item The freezing temperature $T_F(x)$ is an increasing function of the
  stiffness and it does not show to attain any asymptotic finite
  value;
\item At a critical value of the stiffness, $T_F$ becomes higher than
  $T_\th$, and the freezing transition happens without an
  intermediate globular stage.
\end{enumerate}

It is remarkable that recently a phase diagram strikingly 
similar to fig.3 was found in a (variational approximation to a) 
model of random copolymers \cite{MKD}. In this model, the freezing 
temperature shows a power law dependence on the variance $\D\eps$ of 
the monomer pair potentials, $T_F \sim \D\eps^\alpha$, in astonishing 
analogy with our numerical result $T_F \sim x^{0.65}$ (although the 
latter most likely does not give the correct behavior at $x\to 0$).
It is not clear whether this coincidence is fortuitous or has a 
deeper meaning. But in any case it makes the present model more 
interesting as a toy model for protein folding.
The frozen phase of the present model is too ordered to be taken as a 
good model of the native state of a protein (most of the bonds are
parallel), but it is conceivable that the frozen disorder of 
amino acid sequences plays a similar role as the stiffness included 
in the present model.

A more direct application of the present model might be to very 
long semiflexible polymer chains such as DNA or actin. Of course 
such polymers do not live on lattices. Thus they can be deformed 
continuously, while only discrete deformations are possible in our 
model. It is well known that going from a continuous to a discrete 
system can have a big effect on phase transitions. For spin systems, 
the Mermin-Wagner theorem says that this effect is mainly confined to 
2 dimensions, but it is not a priori obvious that the same is true 
in the present case. 

Finally, we have verified that the used algorithm, PERM, is an 
excellent tool for studying polymeric systems at very low energies 
where all other known methods fail. Indeed, we have since used it 
to find ground states in lattice polymer models where low-lying 
states have been given in the literature. In these simulations, 
results of which will be published elsewhere \cite{FGG}, we have been 
able to find putative ground states in {\it all} cases. In several 
of these cases, we also found states lower than these putative ground
states.

\eject

\begin{figure}[ht]
  \centerline{\psfig{file=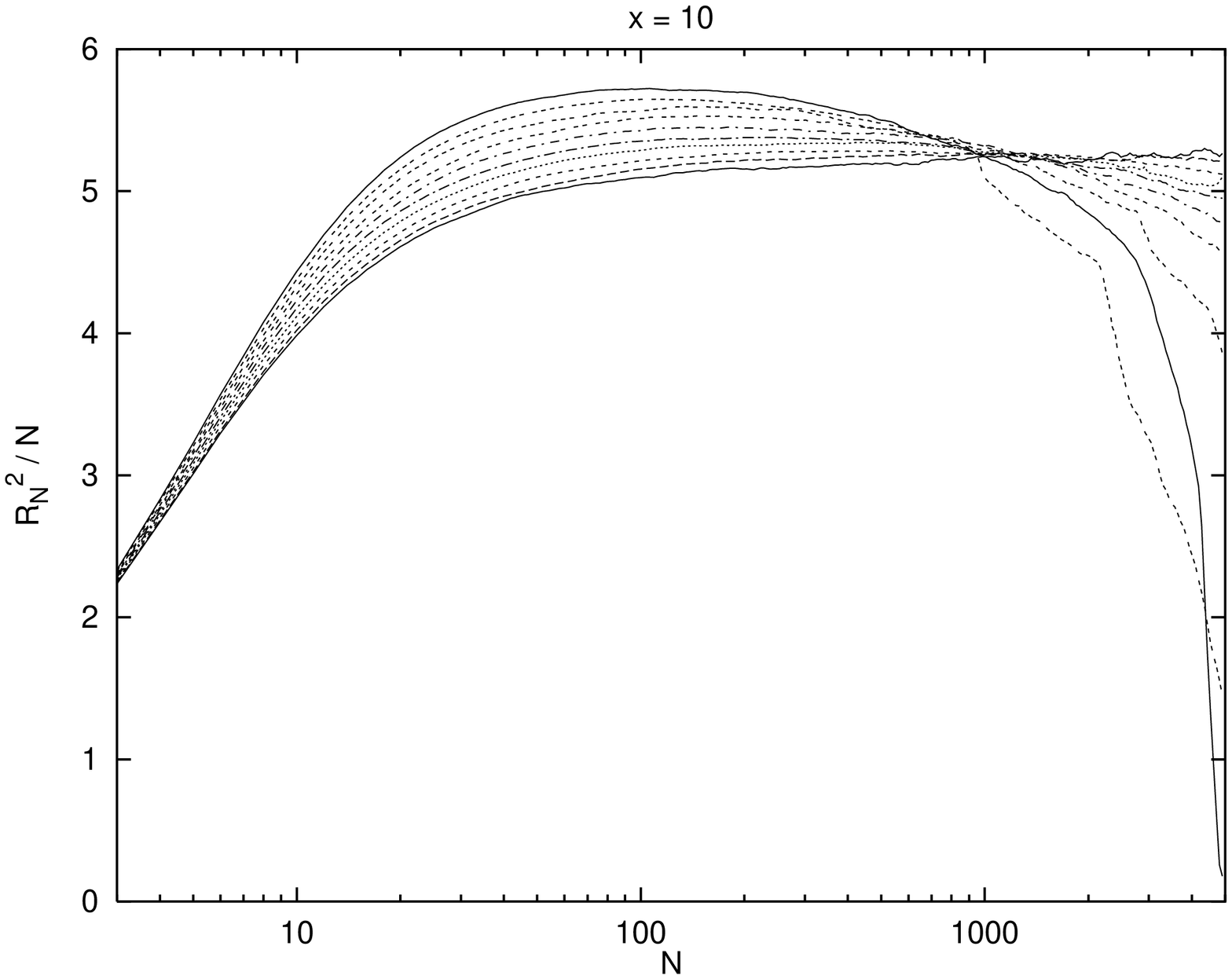,width=12cm,angle=270}}
\caption{\small $R_{e,N}^2/N$ for chains with $x=10$, and for
Boltzmann factors $q=1.247, 1.249, 1.252, 1.255, 1.258, 1.261, 1.264,$
and 1.267 (from flattest to most curved lines)}
\label{Re-10.0}
\end{figure}

\begin{figure}[ht]
  \centerline{\psfig{file=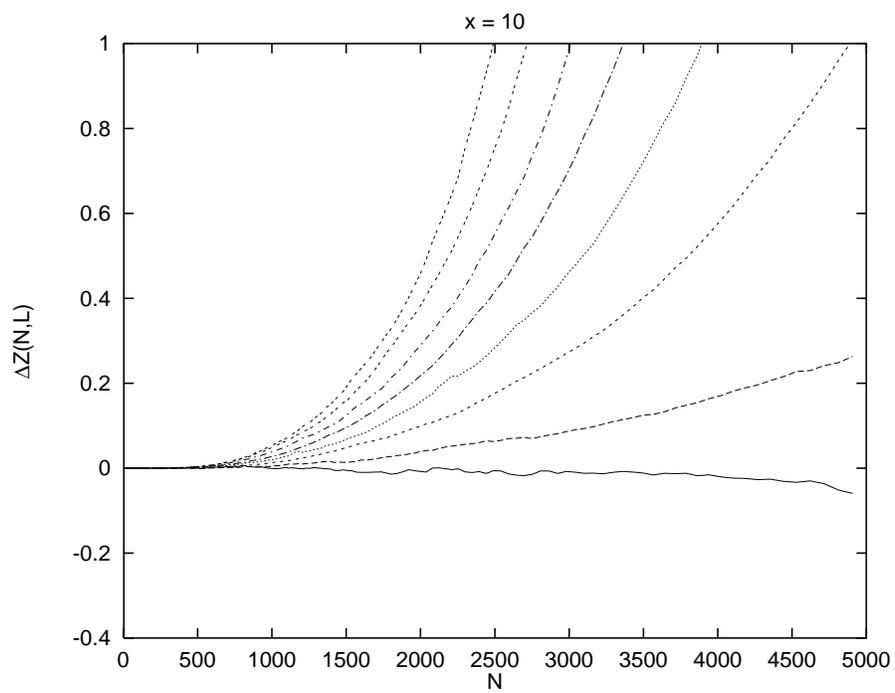,width=12cm,angle=270}}
\caption{\small Partition sum differences for $x=10$ and $L=64$, and 
for the eight highest energies shown also in fig.1 (from bottom to top)}
\label{dZ-10.0}
\end{figure}
 
\begin{figure}
\centering
\epsfysize=12.0cm 
\epsfxsize=15.0cm 
\epsffile{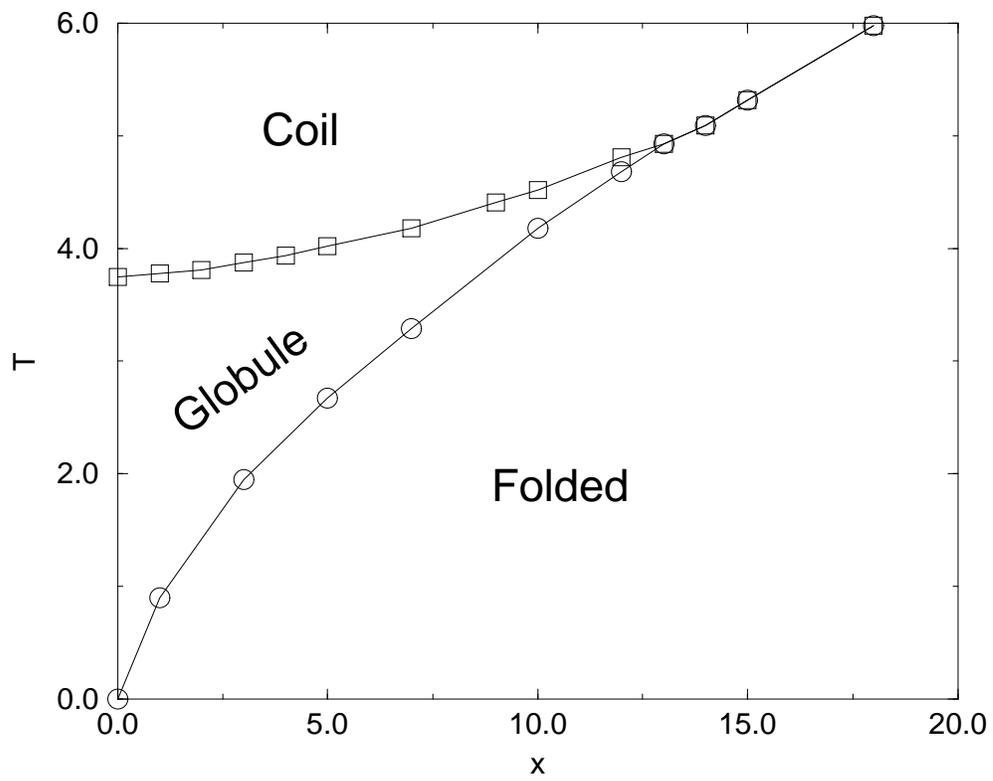}
\caption{\small Phase diagram }
\label{fig_phase}
\end{figure} 

\begin{figure}[ht]
  \centerline{\psfig{file=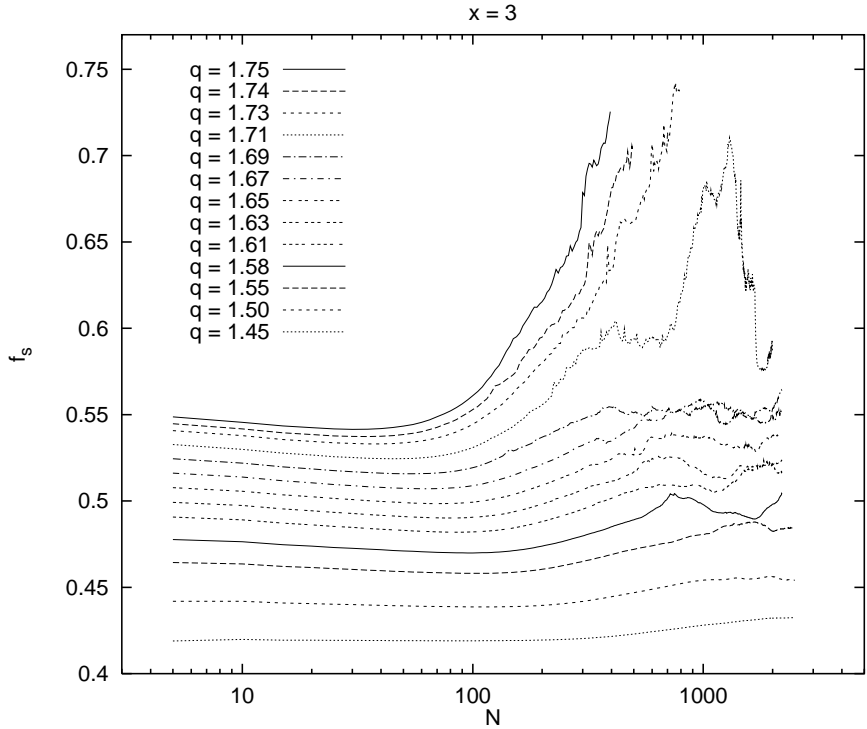,width=12cm,angle=270}}
\caption{\small Fraction $f_s$ of straight joints for $x=3$ and for 
different temperatures, plotted against chain length $N$. At the 
freezing temperature, it should jump from $(1+4/q^x)^{-1}$ to 
a constant $\approx 1$.}
\end{figure}

\begin{figure}[ht]
  \centerline{\psfig{file=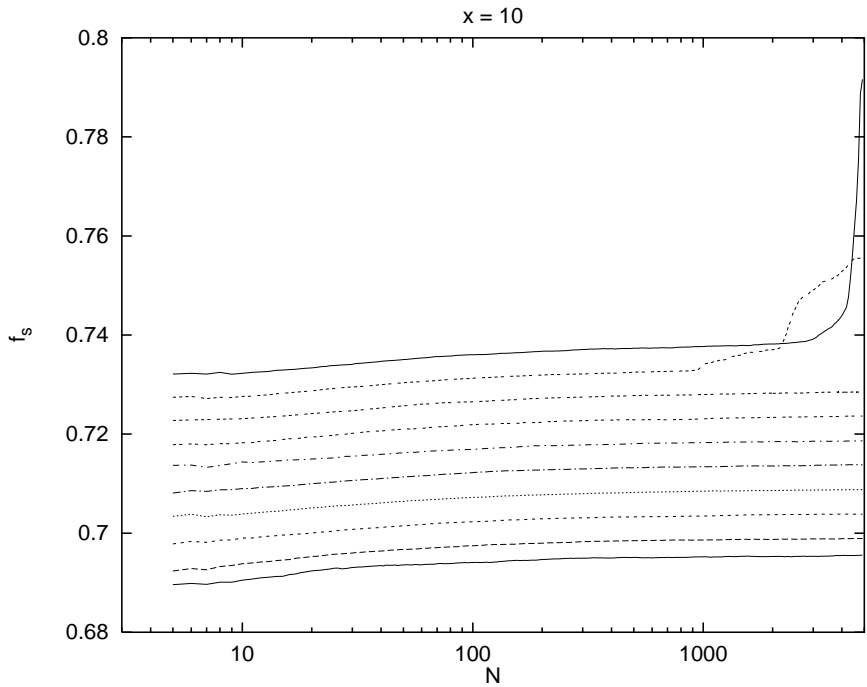,width=12cm,angle=270}}
\caption{\small Same as fig.4, but for $x=10$. The temperatures are 
the same as in fig. 1 (from bottom to top). The freezing transition 
seems to take place near $T_F=4.28\pm 0.02$.}
\end{figure}

\begin{figure}[ht]
  \centerline{\psfig{file=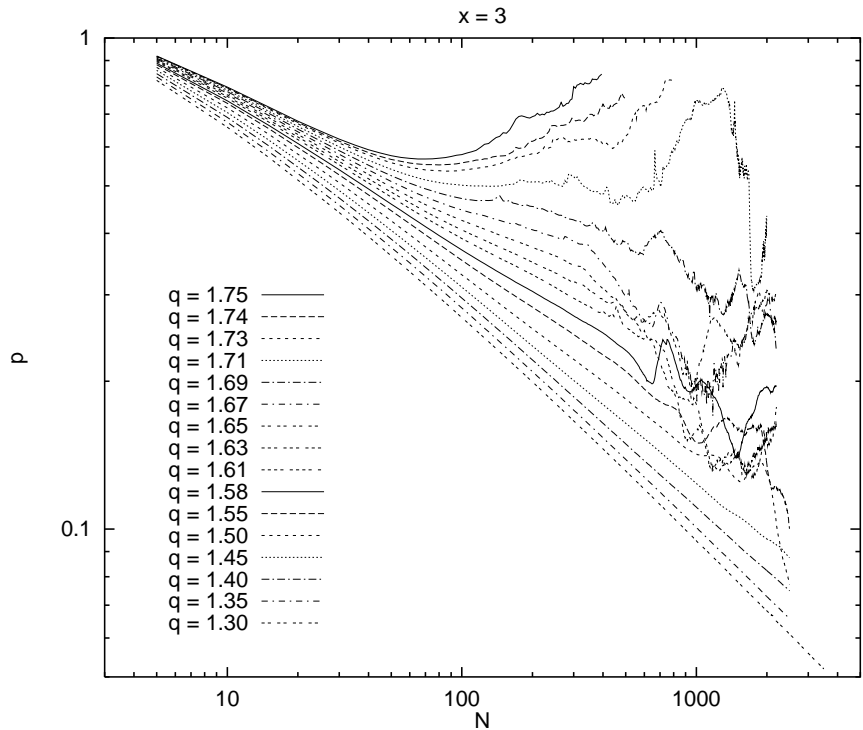,width=12cm,angle=270}}
\caption{\small Anisotropy parameter $p$ for $x=3$ and for the 
same temperatures as in fig.4. At the freezing temperature, its 
behavior should change from power decay to a constant $\approx 1$.}
\end{figure}

\begin{figure}
\centering
\epsfysize=12.0cm
\epsfxsize=15.0cm
\epsffile{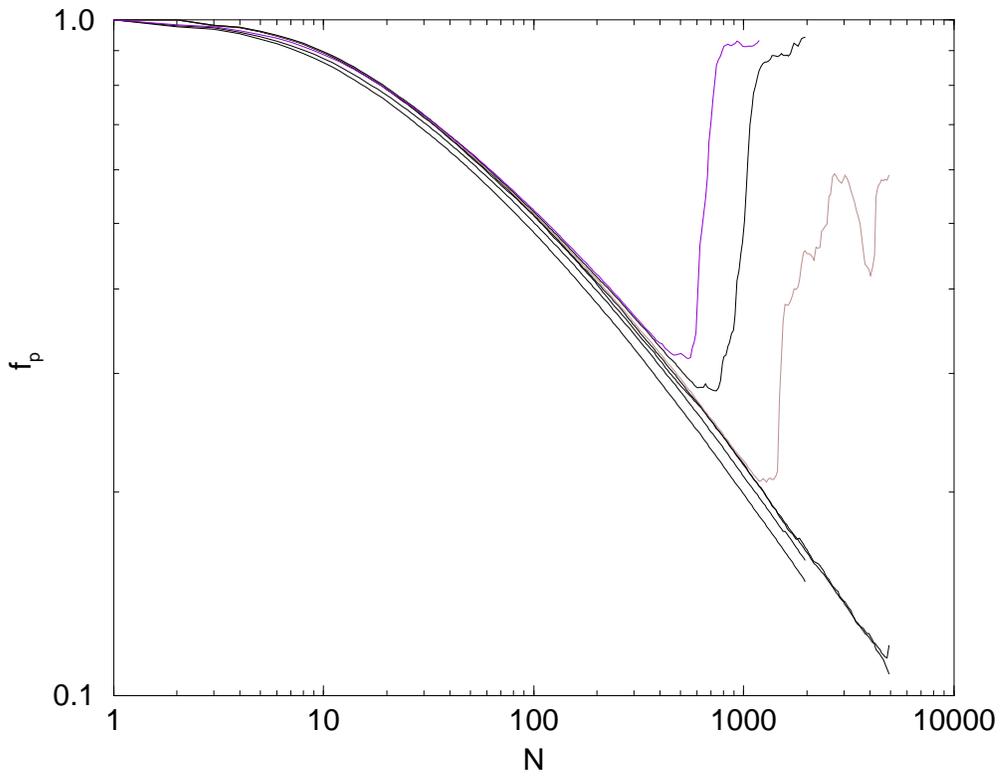}
\caption{\small Same as previous figure, but for $x=15$. At this
stiffness the transition from coil to frozen state takes place
directly. From bottom to top the curves represent $T=5.613, 5.435,
5.3625, 5.339, 5.315, 5.292$, and 5.246. The estimated value of $T_F$
is $5.335\pm 0.01$.}
\label{fig_x15}
\end{figure}
 
\begin{figure}
\centering
\epsfysize=12.0cm
\epsfxsize=15.0cm
\epsffile{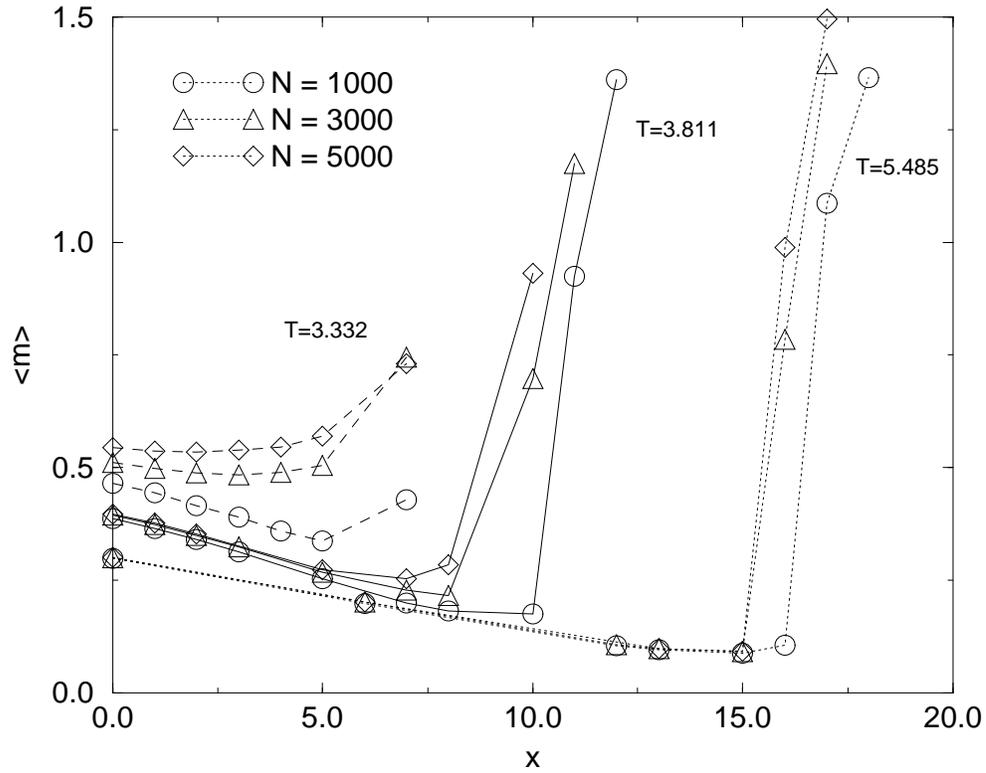}
\caption{\small 
  Average number of contacts $\langle m\rangle$ per monomer as a
  function of the stiffness for fixed temperatures. Three different
  temperatures are considered: $T=3.332$ (below $T_\th$), dashed lines;
  $T=3.811$ (between $T_\th(0)$ and $T_{\rm triple}$), solid lines; 
  and $T=5.485$ (above $T_{\rm triple}$), dotted lines. For each value 
  of $T$ we show data for 3 different chain lengths ($N=$ 1000, 3000, 
  and 5000), since finite size effects are very important. In particular, 
  one sees that minima of $\langle m\rangle$ are reached at 
  $N$--dependent values of $x$, 
  and the effective freezing temperature strongly increases with $N$.}
\label{fig_cont}
\end{figure}
 
\end{document}